\documentclass[preprint,pre,aps]{revtex4}
\usepackage{graphicx}
\usepackage{epstopdf}
\newcommand{\ro}{\hat{\rho}}
\bibstyle{plain}

\begin{document}

\title[]{Self-assembly of lipids in water. Exact results from a one-dimensional lattice model}
\author{Jakub P\c ekalski}
\address{Institute of Physical Chemistry,
Polish Academy of Sciences, 01-224 Warszawa, Poland}
\author{Pawe{\l} Rogowski}
\address{Institute of Physical Chemistry,
Polish Academy of Sciences, 01-224 Warszawa, Poland}

\author{Alina Ciach}
\email{aciach@ichf.edu.pl}
\address{Institute of Physical Chemistry,
Polish Academy of Sciences, 01-224 Warszawa, Poland}

\keywords{self-assembly,lattice models}
\date{\today}

\begin{abstract}
  We consider a lattice model for amphiphiles  in a solvent with molecules chemically similar
  to one  part of the amphiphilic molecule.
  The dependence of the interaction potential on orientation of the amphiphilic molecules is taken into account explicitly.  
 The model is solved exactly
  in one dimension by the transfer-matrix method. In particular, pressure as a function of  concentration,  
  correlation function and  specific heat are calculated.
  The model is compared with the recently introduced lattice model for colloidal self-assembly, 
  where the particles interact with the isotropic
  short-range attraction and long-range repulsion (SALR) potential. 
  Similarities between the amphiphilic and the colloidal self-assembly are highlighted. 
\end{abstract}

\maketitle

\section{Introduction}

Statistical thermodynamics of simple liquids and their mixtures has been extensively studied,
and  thermodynamical and structural properties of such systems are  well understood~\cite{boublik:80:0}. 
In particular, an accurate  equation of state of the Lennard-Jones fluid has been obtained~\cite{kolafa:94:0}. 
The impressive development of the theory was possible thanks to the key contributors including  
prof. Tomas Boublik and prof. Ivo Nezbeda. 
In contrast, the statistical thermodynamics of the  so called soft matter systems is much less developed, and 
recently these systems draw increasing attention.
Complex molecules, nanoparticles, colloid particles or polymers in various solvents interact with effective potentials that may have
quite different forms. When the shape of the effective potential resembles the shape of interactions between 
atoms or simple molecules,
then analogs of the  gas-liquid and liquid-solid transitions occur~\cite{nguyen:13:0}. 
If, however, there are competing tendencies in the interactions,
then instead of the gas-liquid transition or separation of the components, a self-assembly or a microsegregation may be observed
~\cite{stradner:04:0,campbell:05:0,teubner:87:0,kahlweit:86:0,
gompper:94:0,ciach:01:2}. 

The competing interactions can have quite different origin and form. 
One important example of competing interactions is the so called short-range attraction (SA), and long-range repulsion (LR) 
SALR potential~\cite{archer:07:0,archer:07:1,sear:99:0,pini:06:0,ciach:10:1}, consisting of a solvent-induced short-range 
attraction and  long-range repulsion that
is either of electrostatic origin,
or is caused by polymeric brushes bound to the surface of the particles. The attraction favours formation of small clusters. 
Because of the repulsion at large distances, however, 
large clusters are  energetically
unfavourable. For increasing concentration of the particles elongated clusters and a network were observed
in both experiment and theory
~\cite{campbell:05:0,candia:06:0,toledano:09:0,archer:08:0,ciach:10:1}.

Competing interactions of a quite different nature are present in systems containing amphiphilic
molecules such as surfactants, lipids
or diblock copolymers~\cite{ciach:10:1,gompper:94:0}. Amphiphilic molecules are composed of covalently bound polar and organic parts,
and in polar  solvents self-assemble into spherical or elongated micelles, or form a  network in the sponge phase.
In addition, various lyotropic 
liquid crystal phases can be stable~\cite{ciach:01:2,latypova:13:0}. 

Despite of very different origin and shape of the interaction potentials, 
very similar patterns 
occur on the mesoscopic length scale  in the systems 
interacting with the isotropic SALR potential, and in the amphiphilic solutions with strongly anisotropic
interactions~\cite{matsen:96:0,ciach:10:1}.
The particles 
interacting with the SALR potential self-assemble into  spherical
or elongated clusters or form a network, whereas the amphiphiles self-assemble into spherical or elongated micells or form the sponge phase.
The distribution of the clusters or the micelles in
space and the transitions between ordered phases composed of these objects are very similar.

The origin of the universal topology of the phase diagrams in the amphiphilic and SALR systems was
studied in Ref.\cite{ciach:13:0}. 
It has been shown by a systematic coarse-graining procedure that in the case of weak order
the colloidal and the amphiphilic self-assembly can be described by the same Landau-Brazovskii
functional \cite{brazovskii:75:0}. The Landau-Brazovskii functional was first applied to the 
block-copolymers by Leibler 
in 1980 ~\cite{leibler:80:0}. Later functionals of the same type were applied to microemulsions
\cite{ciach:01:2,gompper:94:0,gozdz:96:1}.  
The Landau-Brazovskii -type functional, however, is appropriate only for weak order, 
where the average density and concentration are smooth,
slowly varying functions
on the mesoscopic
length scale. Moreover, in derivation of the functional various assumptions and approximations were made. 
Further approximations are necessary 
in order to obtain solutions for the phase diagram, equation of state and correlation functions. 
Thus, the question of universality of the pattern formation on the mesoscopic length scale, particularly at low temperatures, 
is only partially solved.

We face two types of problems when we want to compare thermodynamic and structural properties in different
self-assembling systems in the framework
of statistical thermodynamics. First, one has to introduce generic models with irrelevant 
microscopic details disregarded. Second,  one has to make approximations to solve the generic models, or perform simulations.
It is not obvious a priori how the assumptions made in construction of the model and the approximations
necessary for obtaining the solutions influence the results. In the case of simulations the simulation box should be commensurate with 
the characteristic size of the inhomogeneities that is to be determined. It is thus important to introduce generic models for different 
types of self-assembly that 
can be solved exactly. Exact solutions can be easily obtained in one-dimensional models, 
but there are no phase transitions in one dimension for temperatures $T>0$. Nevertheless, the ground state (GS) can give important 
information about energetically favorable ordered structures, and pretransitional ordering for $T>0$ can be discussed based
on exact results for the equation of state, correlation function and specific heat.

A generic one-dimensional lattice model for the SALR potential was introduced and solved exactly in Ref.\cite{pekalski:13:0}.
In this model the nearest-neighbors (nn) attract each other, and the third neighbors repel each other. 
It is thus energetically favorable 
to form clusters composed of 3 particles separated by at least 3 empty sites. The GS is governed by 
the repulsion-to-attraction ratio $J$ and by 
the chemical potential of the particles. An interesting property of the GS is strong degeneracy at the coexistence of the ordered
cluster phase with the gas or liquid phases. Due to this degeneracy the
entropy per site does not vanish. The collection of the microstates present for $T=0$ at  the coexistence
between the periodic and the gas or the liquid phases
can be interpreted as a  disordered cluster or bubble phase respectively.
For $T>0$ pseudo phase transitions between the gas and the periodically distributed clusters, 
and  between the periodically distributed clusters and the
dense liquid phases were obtained. The equation of state (EOS) has a characteristic 
shape that is significantly different from the  EOS in simple fluids.

A one dimensional lattice model for amphiphiles in solvents attracting one of the two parts of the
amphiphilic molecule was introduced in Ref.\cite{ciach:14:3}. 
The two models, one for the SALR  and the other one for the 
amphiphilic system, are defined on the same level of coarse-graining, therefore by comparing the exact results we can draw
conclusions on similarities of self-assembly in these systems, and the origin of these similarities. 
In the model introduced in Ref.\cite{ciach:14:3} the GS is strongly degenerate at the phase coexistence between the pure solvent
and periodically distributed bilayers, and the entropy per site does not vanish. Thus, the GS of the 
two models  show remarkable similarity, despite quite different interaction potentials. 
This suggests similar origin of the formation of disordered phases with mesoscopic inhomogeneities in various systems with competing interactions. 

In this work we solve exactly the 
model introduced in Ref.\cite{ciach:14:3} by the transfer matrix method.
We  describe  the model and its  ground state in sec.2. The transfer matrix and exact expressions for the grand potential, density
and correlation function 
are given in sec.3. In sec.4 and 5 we present our results for the EOS and the correlation function. 
In sec.6 we calculate the specific heat with fixed chemical potential in the grand canonical ensemble, and
by using thermodynamic relations we obtain the specific heat with fixed concentration. 
To check our exact calculations we also perform Monte Carlo simulations and compute the specific heat with fixed concentration
directly in the canonical ensemble. Our additional purpose is to verify the results of the simulations
performed in a finite system by comparison with the exact results obtained in the thermodynamic limit.
We cross-check the exact and the simulation
results for the specific heat, because in simulations of the 2- or 3-dimensional self-assembling systems the peak
in the specific heat is interpreted as a signature of a phase transition\cite{imperio:06:0}. 
It is thus worthwhile to compare the simulation and the 
exact results for this important quantity.  In sec.7
our  results for the mechanical, structural and thermal properties are compared with the corresponding results obtained 
in Ref.\cite{pekalski:13:0} for the SALR model. 

\section{The model and its ground state} 
\subsection{The model}

The amphiphilic molecules 
consist of a hydrophilic head and a hydrophobic tail, therefore the interactions between 
them depend on orientations. In the case of a two-component mixture of a polar solvent and amphiphilic molecules, 
for example lipids,
we assume that the solvent molecules attract the
polar head, and effectively repel the hydrophobic tail of the amphiphilic molecule. We  neglect orientational 
 degrees of freedom of the solvent molecules. 
 In a one-dimensional model the continuum of
 different orientations of amphiphiles is reduced to just two  orientations (Fig.1). We assume that the molecules occupy lattice sites, and
 the lattice constant $a$ is of order of the length of the amphiphilic molecule in this model. 
If the solvent molecules are much smaller than the amphiphilic molecules, we assume that the site is occupied by 
a cluster of several solvent molecules.

\begin{figure}
\includegraphics[scale=1]{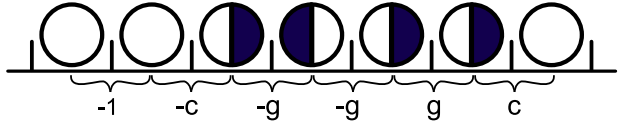}
 \caption{The interacting  pairs of occupied sites in the 1d model. 
 The open circle represents the solvent molecule, and 
the light and dark semicircles represent the head and the tail of the amphiphilic molecule respectively. 
The unit of the inscripted energies is the absolute value of the solvent-solvent interaction energy.}
\label{fig_energies}
\end{figure}

We assume nearest-neighbor interactions.
The absolute value of the energy of two clusters of solvent molecules that occupy the nearest-neighbour sites, $-b$, 
is taken as the energy unit. We assume that the
 interaction between the cluster of solvent molecules and the amphiphilic molecule in the favorable (unfavorable) 
 orientation is $-cb$ ($+cb$), and the
 interaction between two amphiphilic molecules in the favorable and unfavorable  orientation is$-gb$ 
 and $+gb$ respectively.
The  orientations of two amphiphilic molecules are favorable when they are oriented either head-to-head or tail-to-tail.
 The neighborhood of the polar head and the hydrophobic tail is unfavorable. 
The energies of different pairs of occupied sites are shown in Fig. \ref{fig_energies}.
The model is similar to a lattice model of ternary
oil-water-surfactant mixtures introduced in Ref.\cite{ciach:88:0} and to a continuous model of binary mixtures
with amphiphiles~\cite{ciach:11:1}. 

Different values of the parameters $b,c,g$ may correspond to different particular mixtures. In this
work we are interested in general aspects of the amphiphilic self-assembly, especially in similarities
between ordering on the mesoscopic length scale in the amphiphilic and colloidal systems, 
and in origin of these similarities. For this reason we shall not 
try to fit the model parameters to any particular mixture. A representative example of a system described by this
model is a mixture of lipids and water. We should stress that water is a complex 
liquid~\cite{brovchenko:08:0,nezbeda:07:0,nezbeda:11:0,ciach:08:0},
and micro-heterogeneities are present in aqueous solutions
of polar molecules \cite{perera:13:0,nezbeda:87:0}. However, on the mesoscopic length scale of tens or hundreds of nanometers 
the ordering of the water molecules 
plays a subdominant role.


We introduce the microscopic densities $\hat\rho_i(x)$ with $i=1,2,3$ denoting the cluster of solvent molecules, and the
amphiphile with the head on the left
 and on the right respectively.
 $\hat\rho_i(x)=1$ when the site $x$ is in the state $i$ 
and $\hat\rho_i(x)=0$ otherwise. 
  Multiple occupancy of the lattice sites is excluded.
   We further restrict our attention to the liquid phase and assume close-packing,
\begin{equation}
\label{1}
 \sum_{i=1}^3\hat\rho_i(x)=1.
\end{equation}

Up to a state-independent constant the Hamiltonian of an open system, with the chemical-potential contribution included, 
can be written in the form
\begin{equation}
 H[\{\hat\rho_i\}]/b=
 \frac{1}{2}\sum_{x=1}^L\sum_{x'=1}^L\hat\rho_i(x)V_{ij}(x,x')\hat\rho_j(x')-\sum_{x=1}^L\mu\hat\rho_1(x),
\end{equation}
where the  summation convention for repeated indices is used, $L$ is the system size,    $\mu=\mu_1-\mu_s$, 
$\mu_1b$ is the chemical potential
 of the solvent, and  the chemical  potential of amphiphiles,  $\mu_sb=\mu_2b=\mu_3b$,
 is independent of orientations of the molecules. We assume periodic boundary conditions, $L+1\equiv 0$.
 According to the above discussion of interactions the interaction potential ${\bf V}$ is
\begin{eqnarray}
{\bf V}(x,x+1) = \left[
\begin{array}{rrr}
 -1\;& -c\;&  c\\
c\;& g\;& -g\\
-c\;& -g\;& g
\end{array}
\right]
\label{bfV}
\end{eqnarray}
and $V_{ij}(x,x-1)=V_{ji}(x,x+1)$, $V_{ij}(x,x+k) = 0$ for $|k|>1$. 
In the liquid phase we can neglect density fluctuations (Eq.(\ref{1})), hence $\hat\rho_1(x)=1-\hat\rho_2(x)-\hat\rho_3(x)$
and there are two independent densities. 
In a disordered phase $\langle \hat\rho_i(x)\rangle=\rho_i$, and $\rho_1=1-\rho_s$ with $\rho_s=2\rho_2=2\rho_3$ denoting the 
average amphiphile concentration. 
\subsection{The ground state}

At $T=0$ the stable structure corresponds to the global minimum of the Hamiltonian. 
 Apart from the solvent-rich and amphiphile-rich phases we find
 stability of the  periodic phase where amphiphilic bilayers are separated by  layers of solvent.
 In the amphiphile-rich phase the molecules are oriented head-to-head and tail-to-tail when $g>0$. 
In the periodic phase a solvent-occupied site is followed by $2$ sites occupied by
properly oriented amphiphilic molecules.
  The coexistence lines obtained by equating  $H/L$ in these phases are \cite{ciach:14:3}
\begin{eqnarray}
\label{GSceox}
 \mu=\left\{ \begin{array}{ll}
g-1 &\quad \textrm{solvent-amphiphile}\\
\frac{2c+g-3}{2} &\quad \textrm{solvent-bilayers}\\
2(g-c) &\quad \textrm{bilayers-amphiphile}.\\
                 \end{array}\right.
\end{eqnarray}

\begin{figure}
\includegraphics[scale=1]{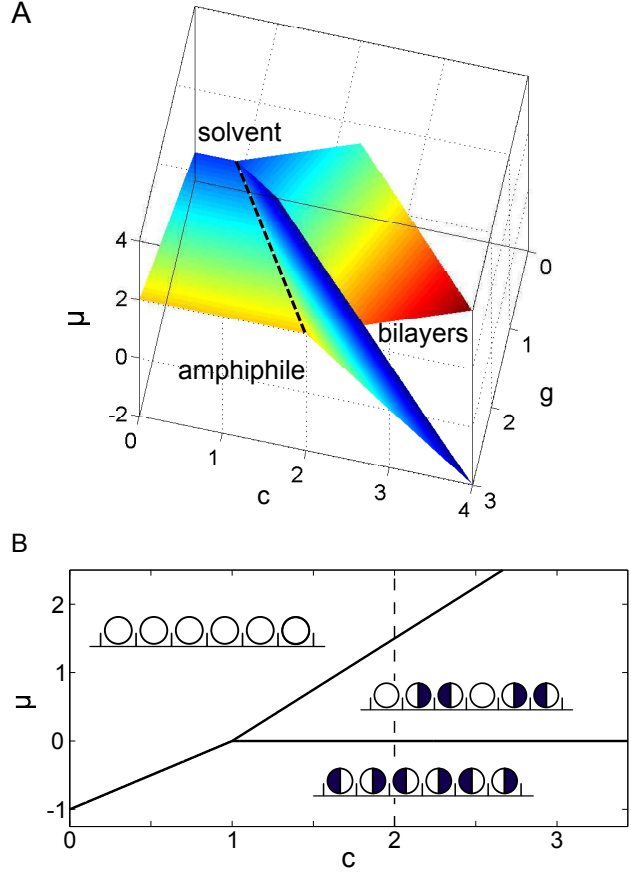}
 \caption{The ground state of the model in the variables $(c,g,\mu)$  (panel A) and in the plane $(c,\mu)$ for $c=g$ (panel B). 
  The surfaces in panel A represent coexistence between the phases,
while the black dashed line represents the triple points where all the three phases coexist.
The triple point for $g=c$ is at  $(c,\mu)=(1,0)$. 
On the B panel a schematic illustration of the three phases is shown in the insets inside the region of 
stability of each phase. 
At the coexistence with the solvent-rich phase the layers of solvent, $l$, can have an arbitrary length, 
and at the coexistence
with the amphiphile-rich phase the layers of amphiphilic molecules can have the thickness $2n$ with arbitrary $n$.
The dashed line corresponds to $c=g=2$ for which the EOS, the correlation function and the specific heat are calculated 
in the following sections.}
 \label{fig_gs}
\end{figure}

The $(c,g,\mu)$ ground state is shown in Fig. \ref{fig_gs}.
The solvent-amphiphile coexistence occurs for small values of $c$. All the three phases coexist at the triple line
$g=2c-1$ and $\mu=2(c-1)$. 
When $g\le 2c-1$ the periodic structure of solvent-separated bilayers may be present for some range of $\mu$ and
 the sequence of the
 stable phases for increasing $\mu$ at fixed  $c$ and $g$ is: amphiphiles-bilayers-solvent.

At the solvent-periodic phase coexistence the 
separation $l$ between the  bilayers can be arbitrary, because $H/L$ is independent of $l$ for $\mu$ given
by Eq.(\ref{GSceox})b~\cite{ciach:14:3}. 
Thus, the ground state is strongly degenerate and the entropy per site does not vanish.
 Similar degeneracy occurs at the periodic-amphiphile phase coexistence. 
 At the periodic-amphiphile coexistence  the separation between the solvent occupied  sites is  $2n$ with arbitrary $n$,
 because when Eq.(\ref{GSceox})c holds, $H/L$ is independent of $n$~\cite{ciach:14:3}.

Note that the arbitrary separation between the bilayers at the 
coexistence between the solvent and the periodic phase signals
 vanishing surface tension between the two phases. 
 Similar degeneracy of the ground state was found earlier for the lattice model of microemulsion \cite{ciach:89:0}.
 The very low surface tension at the coexistence between the 
microemulsion and the water-rich phases was attributed to the amphiphilic nature of surfactant molecules.
 
\section{Exact expressions for pressure, density and the correlation function in terms of the transfer matrix}
In this section we introduce the transfer matrix and develop exact expressions for the grand thermodynamic potential, the average
density of each component and for the correlation function.
The elements of the transfer matrix ${\bf T}$ are given by
\begin{equation}
{\bf T}_{ij} \equiv \exp \left\{-\beta^* \left(V_{ij} - \mu \delta_{i1}^{Kr} \right)\right\},
\end{equation}
where $\delta_{i1}^{Kr}$ is 1 for $i=1$ and 0 otherwise, and $\beta^*=1/T^*=b/(k_BT)$ with $k_B$ denoting the Boltzmann constant.
The partition function for a system with periodic boundary conditions is
\begin{equation}
\Xi = \sum_{\ro(1)} \ldots \sum_{\ro(L)} \prod_{n=1}^{L} \exp \left\{ -\beta^* \left[ \ro^T(n) {\bf V} \ro(n+1)
- \mu \ro_1(n) \right] \right\}.
\end{equation}
where  $\hat\rho(x)^T=(\hat\rho_1(x),\hat\rho_2(x),\hat\rho_3(x))$ denotes the microscopic state at the site $x$, and
is  transverse to the columnar vector $\hat\rho(x)$. At each lattice site there can be one of the 3 microscopic states $(1,0,0)$, $(0,1,0)$, or $(0,0,1)$.
In terms of the transfer matrix $\Xi$  takes the form
\begin{equation}
\Xi = \textrm{Tr} {\bf T}^L = \lambda_1^L + \lambda_2^L + \lambda_3^L,
\end{equation}
where $\lambda_i$ is the eigenvalue of the transfer matrix. If we denote $\lambda_1 = max_{i\in \{1,2,3\}} ( |  \lambda_i|)$,
the partition function for the system size $L\gg 1$
 takes the even simpler asymptotic form
\begin{equation}
\Xi \simeq \lambda_1^L.
\end{equation}
In the thermodynamic limit  the grand  potential in $b$ units, $\Omega^*=\Omega/b$,  is given by the exact formula
\begin{equation}
\label{Omega}
\lim _{L\to\infty}\Omega^*/L =-p^*=- T^*\lim _{L\to\infty} \frac{\ln \Xi}{L}= -T^* \ln \lambda_1,
\end{equation}
where $p^*$ is 1 dim. pressure in $b/a$ units. 

The average density of the $i$-th state  is independent of $x$ because of the translational invariance, and is given by
\begin{eqnarray}
\langle \ro_i(1) \rangle &=& \frac{1}{\Xi} \sum_{\ro(1)} \ldots \sum_{\ro(L)} \prod_{n=1}^{L} \ro_i(1) 
\exp \left\{ -\beta^* \left[ \ro^T(n) {\bf V} \ro(n+1) - \mu \ro_1(n) \right] \right\}.
\end{eqnarray}
If we change the basis of {\bf T} with a help of the invertible matrix {\bf P} such that $ \bf P^{-1} T P$ is diagonal, 
then the average density in thermodynamic limit is given by the following expression:
\begin{equation}
\label{avr}
\langle \ro_i \rangle  = \langle \ro_i(1) \rangle = \lim_{L\to \infty}\frac{1}{\Xi} {\bf P}_{1i}^{-1} \lambda_1^L {\bf P}_{i1} 
= {\bf P}_{1i}^{-1} {\bf P}_{i1}.
\end{equation}

The correlation function $G_{ii}(n)$ between two sites in the same state $i$ separated by $n$ sites is given by
\begin{equation}
\label{G}
G_{ii}(n) = \langle \ro_i(1) \ro_i(n+1) \rangle - \langle \ro_i(1)\rangle \langle \ro_i(n+1) \rangle,
\end{equation}
where
\begin{eqnarray}
\langle \ro_i(1) \ro_i(n+1) \rangle =  \frac{1}{\Xi}  {\bf T}^{n}[\ro_i(1),\ro_i(n+1)] {\bf T}^{L-n}[\ro_i(n+1)\ro_i(1)].
\end{eqnarray}
We change  the basis to the one in which {\bf T} is diagonal, take the thermodynamic limit and 
obtain the exact formula,
\begin{eqnarray}
\label{rr}
<\ro_i(1) \ro_i(n+1)>&=&   \sum_{k = 1}^{3} \Big( \frac{\lambda_k}{\lambda_{1}} \Big)^{n} 
\mathbf{P}_{ik}  {\mathbf{P}^{-1}}_{ki}\mathbf{P}_{i1} \mathbf{P}_{1i}^{-1} \nonumber \\
&=&   < \ro_i>^2   + \sum_{k = 2}^{3}  \Big( \frac{\lambda_k}{\lambda_{1}} \Big)^{n} A^{(k)}_{ i}  B^{(k)}_{i},
\end{eqnarray}
where 
\begin{equation}
A^{(k)}_{ i} =  \mathbf{P}_{ik}\mathbf{P}_{1i}^{-1}, \quad B^{(k)}_{i} = \mathbf{P}^{-1}_{ki}\mathbf{P}_{i1}
\end{equation}
From (\ref{rr}) and  (\ref{G}) we obtain the correlation function 
\begin{equation}
G_{ii}(n) = \sum_{k = 2}^{3}  \Big( \frac{\lambda_k}{\lambda_1} \Big)^{n}  A^{(k)}_{i}  B^{(k)}_{ i} .
\label{corr_func}
\end{equation}
Eq. (\ref{corr_func}) can be further simplified for $ n \gg 1 $. 
In such a case we can neglect the smallest components of the sum in  Eq.(\ref{corr_func}). If the second largest 
(in the absolute value) eigenvalue $\lambda_{2}$ is a pure real number, 
then $G_{ii}(n) = \Big( \frac{\lambda_2}{\lambda_1} \Big)^{n}  A^{(2)}_{i}  B^{(2)}_{ i}  $, 
but if the $\lambda_{2}$ has a non-zero imaginary part, then we have to take into account also the eigenvalue $\lambda_3$, complex 
conjugate to $\lambda_2$. Let us introduce the notations  
\begin{eqnarray}
 \lambda_{2} = Z_{\lambda} e^{\lambda i}\hskip1cm A^{(2)}_{ 1} = Z_{\alpha} e^{\alpha i}\hskip1cm B^{(2)}_{1} = Z_{\gamma} e^{\gamma i}
\end{eqnarray}
 and  
\begin{eqnarray}
\label{xi}
\xi = \Big( \ln(   \frac{\lambda_{1}}{Z_{\lambda}}) \Big)^{-1}.
\end{eqnarray}
In terms of these parameters the correlation function for large separations between the particles in an infinite system 
takes the asymptotic form
\begin{eqnarray}
G_{11}(n) &\stackrel{n \gg 1}{\simeq} &  2  Z_{\alpha} Z_{\gamma} e^{-n/\xi } \cos \Big( n\lambda 
+ \alpha + \gamma \Big).
\label{corr_func_ag}
\end{eqnarray}

\section{Equation of state}

\begin{figure}[th]
\includegraphics[scale=1]{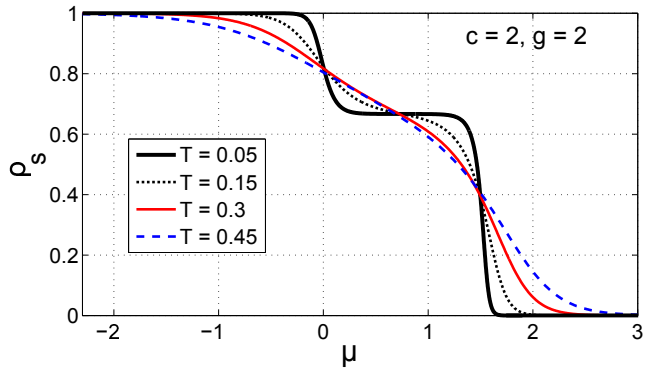}
\caption{Average density of the amphiphile, $\rho_s = 1 - \rho_1 = \rho_2+\rho_3$, for $c=g=2$ at $T^* = 0.05,0.15,0.3,0.45$
as a function of the reduced chemical potential difference $\mu$.
}
\label{eosm}
\end{figure}

  We choose stronger interactions 
between the amphiphilic than between the solvent molecules, $c=g=2$. 
For such interaction parameters the periodic phase is present on the GS (see the dashed line in Fig.2B). 
In Fig.\ref{eosm} we show the concentration $\rho_s$ (average density of the amphiphile) as a function of 
the reduced chemical potential difference $\mu$.
Note the rounded steps for $\mu\approx 0$ and $\mu\approx 1.5$. The steps occur for 
the values of $\mu$  corresponding to the 
GS phase transitions between the periodic and the amphiphile-rich or solvent-rich phases respectively. 
Between the steps  the plateaus for the
three densities, $\rho_s=1,2/3,0$ occur. For increasing $T^*$ the $\rho_s(\mu)$ lines become smoother, but the inflection points 
exist up to $T^*\approx 0.2$.

\begin{figure}[th]
\includegraphics[scale=1]{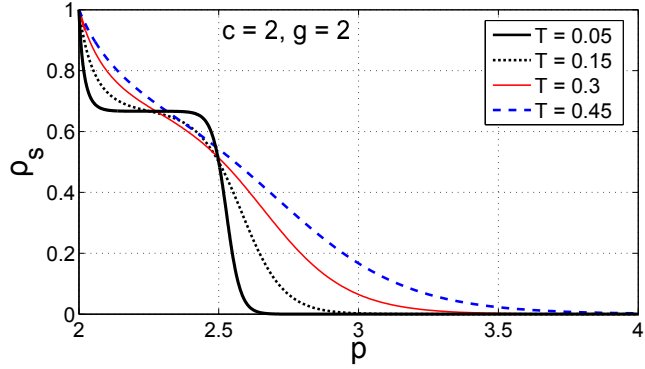}
\caption{Average density of the amphiphile, $\rho_s = 1 - \rho_1 = \rho_2+\rho_3$, for $c=g=2$ at $T^* = 0.05,0.15,0.3,0.45$.
As the energy unit we choose the interaction $b$ between the two solvent molecules, and 
the length unit is the lattice constant $a$.  We assume that $a$ is 
of order of the size of the amphiphilic molecules. In the case of lipids  $a\sim 2nm$.}
\label{eos}
\end{figure}

\begin{figure}[th]
\includegraphics[scale=1] {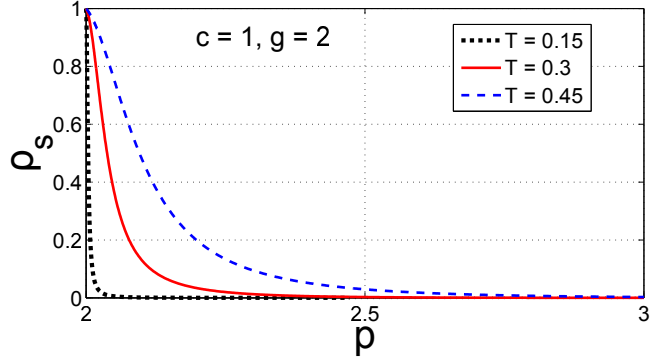}
\caption{Average density of the amphiphile, $\rho_s = 1 - \rho_1 = \rho_2+\rho_3$, for $c=1$ $g=2$ and $T^* =0.15,0.3,0.45$.
As the energy unit we choose the interaction $b$ between the two solvent molecules, and 
the length unit is the lattice constant $a$.  We assume that $a$ is 
of order of the size of the amphiphilic molecules. In the case of lipids  $a\sim 2nm$.}
\label{cgsmall}
\end{figure}

In Eqs.(\ref{Omega}) and (\ref{avr}) the  pressure  and the average density are expressed in terms
of $T^*$ and the reduced chemical potential difference $\mu$. By eliminating $\mu$ we can obtain the dependence of the
amphiphile density
on $p^*$.  We  present $\rho_s(p^*)$ in Fig.\ref{eos}. 
Note that although there are no phase transitions in the strict sense in 1d, for low $T^*$ there is a rapid change in $\rho_s$
between $\rho_s=1$ and $\rho_s=2/3$ for a very small $p^*$ interval near $p^*\approx 2$, 
almost constant amphiphile density between $p^*\approx 2$ 
and $p^*\approx 2.5$, and again a rapid change of $\rho_s$ between $2/3$ and nearly $0$ for $p^*\approx 2.5$. This behavior
suggests the pseudo-phase transitions between the amphiphile-rich and the periodic pseudo-phase with the density $\rho_s\approx 2/3$ 
(see Fig.2b) and next between the periodic 
pseudo-phase and the solvent-rich pseudo-phase. For increasing temperature the changes of the 
slope of the $\rho_s(p^*)$ line for increasing $p^*$ become smaller.
This result should be contrasted with the pressure-concentration dependence shown in Fig.\ref{cgsmall}
for the interaction parameters such that the periodic phase is not stable
at the GS.
\section{Correlation function}
\begin{figure}[th]
\includegraphics[scale=1] {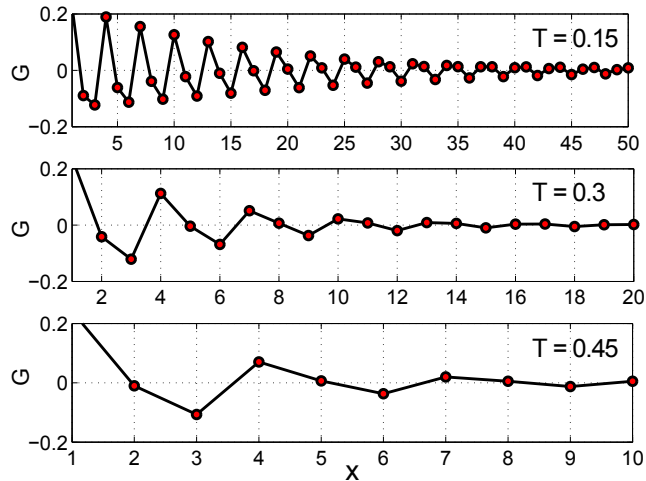}
\caption{Correlation function between the solvent molecules at $T = 0.15, 0.3, 0.45$ for $g=c=2$ and $\mu = 1$.}
\label{cfm}
\end{figure}

In the case of the periodic boundary conditions the system is translationally invariant, 
and the assembly into bilayers should
be reflected in the shape of the correlation function. When the bilayers are formed, then the correlation function for the solvent,
$G_{11}(x)$ should be negative for two subsequent values of $x$, where the properly oriented 
amphiphilic molecules should appear with larger probability than the solvent molecule.
The exact results for $G_{11}(x)$, given in Eq.(\ref{corr_func}), are shown in Fig.\ref{cfm} for
$\mu=1$, i.e. for the GS stability of the periodic phase, for a few rather high temperatures. 
We can see the oscillatory decay of correlations, with the period 3 as in the case of the concentration in the GS periodic phase.
The decay length decreases
with increasing temperature. Only for short distances $G_{11}(x)<0$ for two subsequent values of $x$, however. 

\begin{figure}[th]
\includegraphics[scale=1] {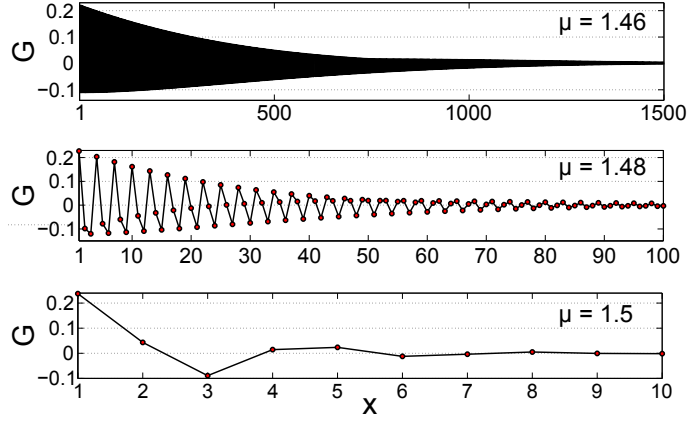}
\caption{Correlation function between  the solvent molecules at $T = 0.005$ for $g=c=2$ and $\mu = 1.46$ (a), $\mu = 1.48$ (b),
$\mu = 1.5$ (c), with $\mu = 1.5$ being the value of the chemical potential at the coexistence between 
the periodic and the solvent-rich
phases at the GS.}
\label{cfT1}
\end{figure}

\begin{figure}[th]
\includegraphics[scale=1] {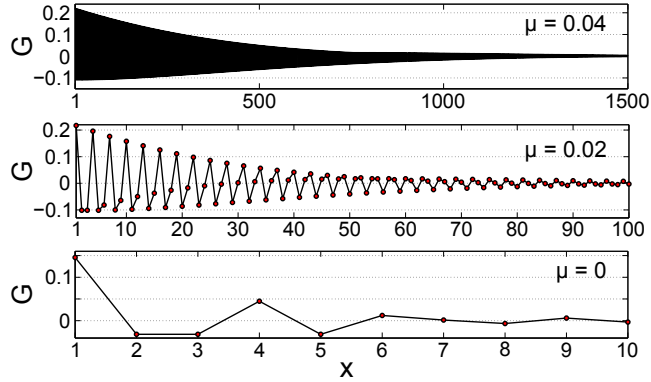}
\caption{Correlation function between  the solvent molecules at $T = 0.005$ for $g=c=2$ and $\mu = 0.04$ (a), $\mu = 0.02$ (b), 
$\mu = 0$ 
(c), with $\mu = 0$ being the value of the chemical potential at the coexistence between the periodic and 
the amphiphile-rich phases at the GS.}
\label{cfT2}
\end{figure}

In Figs.\ref{cfT1} and \ref{cfT2} we show $G_{11}(x)$ for very low $T^*$ and a few values of $\mu$ close to
the GS coexistence between the periodic and the solvent- or amphiphile-rich phases. Very large correlation length, 
3 orders of magnitude larger than the molecular size, can be seen inside the GS stability of the periodic phase.

\begin{figure}[!h]
\includegraphics[scale=1] {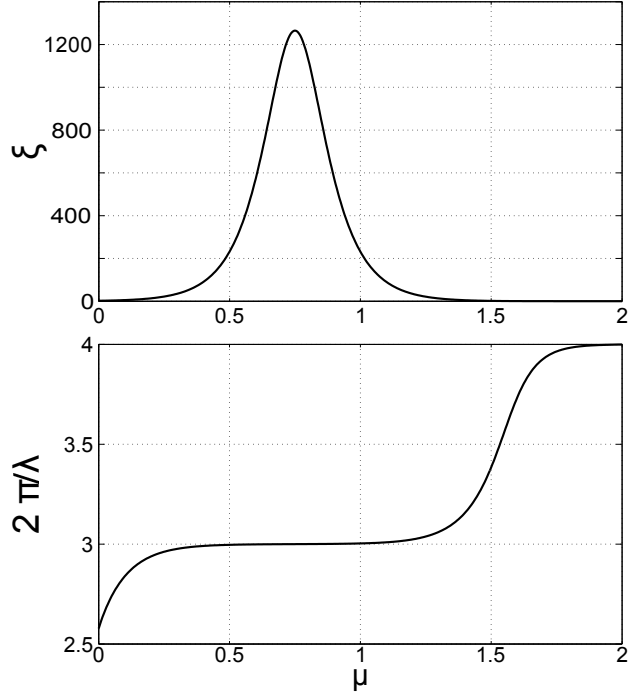}
\caption{The correlation length $\xi$ 
 and the period of the oscillatory decay of correlations (Eqs.(\ref{xi}) and (\ref{corr_func_ag})) 
 for $g=c=2$ and $T = 0.1$. For $\mu = 0$ and $\mu=1.5$ 
 the periodic phase coexists with the amphiphile and the solvent  respectively at $T=0$ .}
\label{xi_l}
\end{figure}

The correlation length $\xi$ (see Eq.(\ref{xi})) and the period of  the damped oscillatory decay (see Eq.(\ref{corr_func_ag})) 
are shown in Fig.\ref{xi_l}. We can see that $\xi\to 0$ beyond the GS stability of the periodic phase, i.e.
for $\mu<0$ and $\mu>1.5$. Moreover,
for $0<\mu<1.5$ the period of the  damped oscillations is $2\pi/\lambda\approx 3$.


 \begin{figure}[!h]
\includegraphics[scale=1] {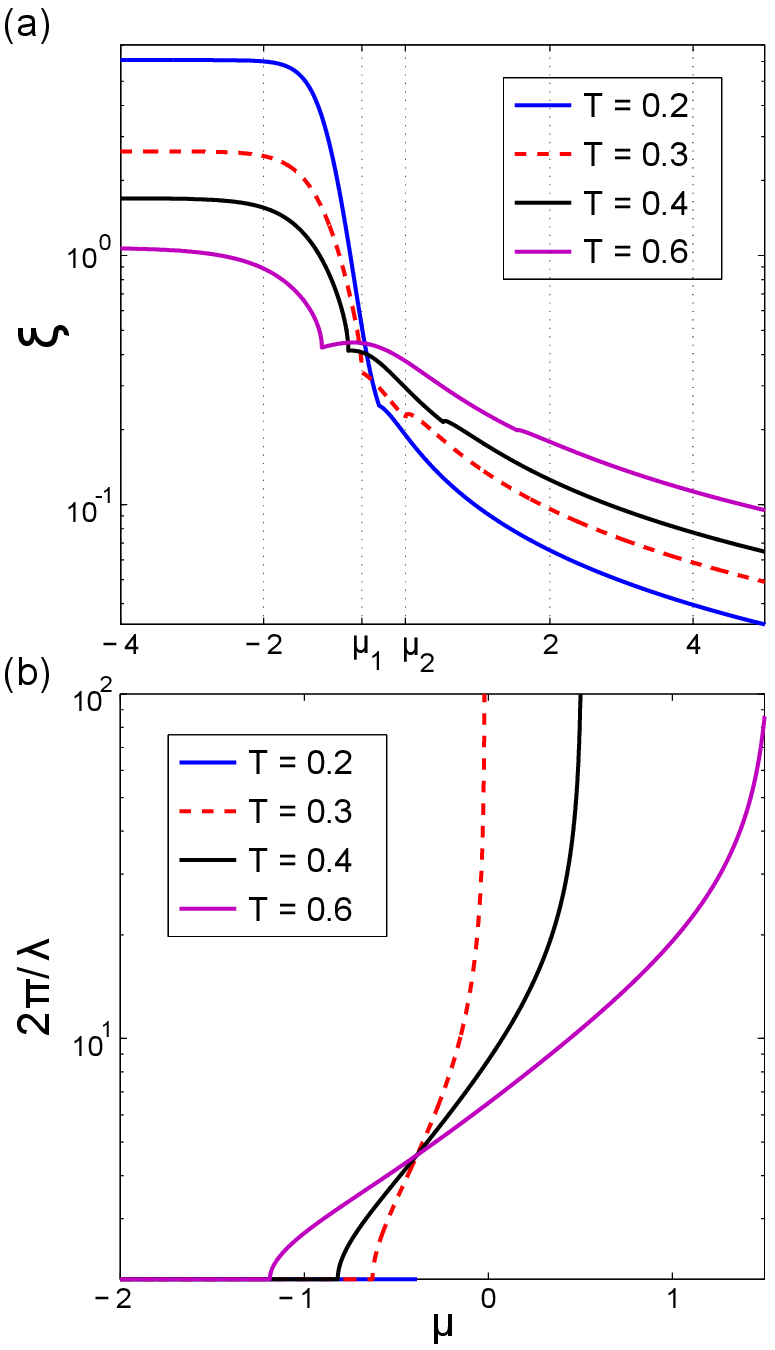}
\caption{The correlation length $\xi$  and the period of the oscillatory decay of correlations for 
$c=0.623 ,g=0.25$ at $T^* = 0.2,0.3,0.4,0.6$.  For $\mu = -0.75$  the amphiphile and the solvent coexist at $T^*=0$.
 For $T^*<0.2$ the monotonic decay of correlations between the solvent molecules is obtained. 
 At $T^*=0.2$ the period of the oscillatory decay jumps from zero to infinity for $\mu \approx -0.39$. For $T^*>0.2$ the
 correlation function for the solvent exhibits an oscillatory decay for some range of $\mu$. In the case of $T^*=0.3$
 the oscillatory decay occurs for $\mu_1<\mu<\mu_2$.
Note the logarithmic scale on the vertical axis.}
\label{xi_l_nn}
\end{figure}

Let us focus on the structure for the interaction parameters corresponding to the absence of the periodically 
distributed bilayers in the GS. We choose  $c=0.623$ and $g=0.25$.
The correlation length $\xi$ and the period $2\pi/\lambda$ of  the damped oscillatory decay 
for $c=0.623$ and $g=0.25$, and different temperatures are shown 
in Fig.\ref{xi_l_nn}. 
 For $T^* \le 0.2$ the eigenvalue  $\lambda_2$ is a pure real number for any value of the chemical potential, and 
 the correlation function decays monotonically.
The change of the $\xi$ slope 
 around $\mu = -0.39$ at $T^*= 0.2$ indicates the point where $\lambda_2$ changes sign.
 The period of the oscillatory decay jumps from zero to infinity at this point. 
 Both cases correspond to the monotonic decay of correlations, but at this point ($\mu = -0.39$ and $T^*= 0.2$)
 an oscillatory decay  for $T^*>0.2$ and a range of  $\mu$ around $\mu = -0.39$ begins. This is because
 for $T^*>0.2$ there are two discontinuities of the derivative of $\xi$ (e.g. at $\mu_1$ and $\mu_2$ for $T^*=0.3$).
 Between the two points of discontinuity of $\partial\xi/\partial\mu$, the $\lambda_2$ is a complex number,
   and hence for this interval of $\mu$ the correlation function has an oscillatory decay. 
   Note that for $c=0.623,g=0.25$ the periodic phase is not stable on the GS, and
   counterintuitively the oscillatory decay of $G_{11}(x)$ occurs at higher $T^*$. 
  However, the correlation length is smaller than the period of the damped oscillations.
  Similar change from the monotonic
 to  the oscillatory decay of correlations for increasing temperature has been obtained for the SALR system 
 in Ref.\cite{pekalski:13:0}.
 
 \section{Specific heat}
 In this section we consider the specific heat for a fixed concentration. Fixed concentration imposes a global
 constraint on the microstates, therefore exact analytical calculations directly in the canonical ensemble are less easy. 
 To overcome this difficulty, we first calculate
 the specific heat for fixed chemical potential  using the exact results of sec.3. 
 In order to compute the  specific heat for fixed 
 concentration we use the thermodynamic relations. For comparison we 
 perform Monte Carlo (MC) simulations in the canonical ensemble.

\begin{figure}[th]
\includegraphics[scale=0.9] {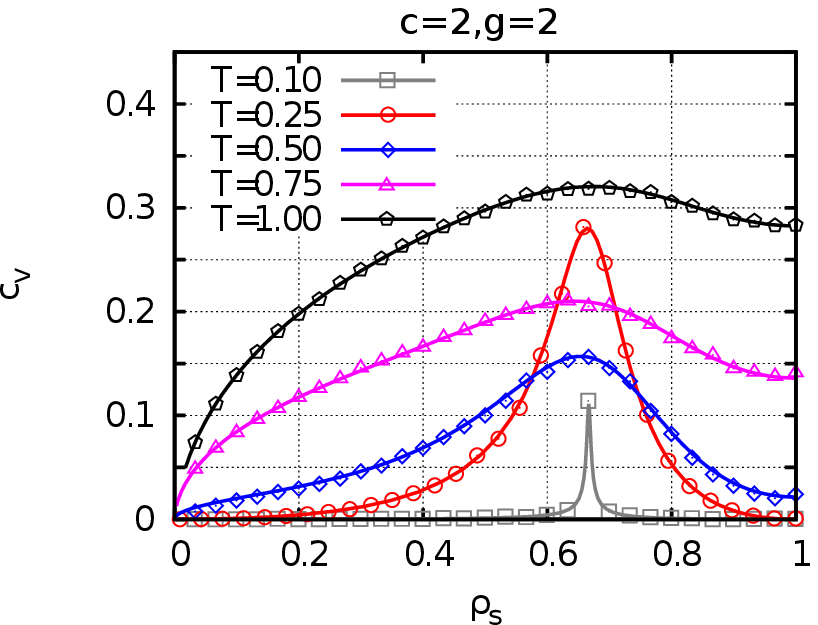}
\includegraphics[scale=0.9] {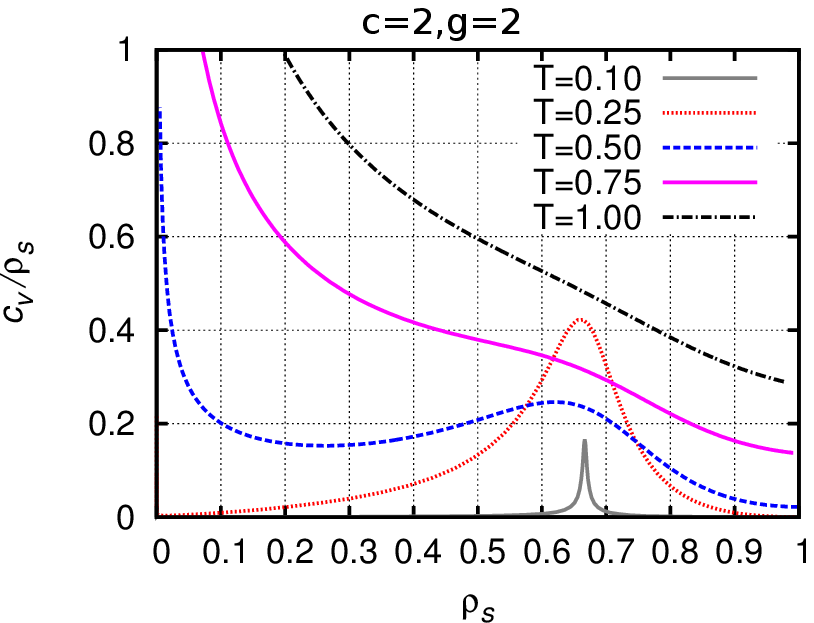}
\caption{ The specific heat for fixed concentration  $\rho_s$ as a function of  $\rho_s$
for $c=g=2$ and  different temperatures.
 Top: per unit volume ($c_v$). Bottom:  per amphiphilic molecule ($c_v/\rho_s$).
 Symbols denote the results of the Monte Carlo simulations for $L=4800$, 
and lines represent the 
 analytical results  in thermodynamic limit (Eqs. (\ref{cmu})- (\ref{rho_mu}) and (\ref{Omega})).
 }
\label{cv22}
\end{figure}

 The specific heat of a mixture  with fixed number of all particles and fixed chemical potential difference
 between the components, $\mu$, is given by
\begin{equation}
\label{cmu}
c_{\mu} = -\frac{T}{L}\left(\frac{\partial^2\Omega^{*}}{\partial T^2}\right)_{\mu,V}
\end{equation}
where the exact expression for $\Omega^{*}$ is given in Eq.(\ref{Omega}). We 
  compute  $c_v$ using the following relations,
\begin{equation}
\label{cv}
c_{v} = c_{\mu}-T\left(\frac{\partial \rho_{\mu}}{\partial T}\right)^2 \left(\frac{\partial \rho_{\mu}}{\partial \mu}\right)^{-1}
\end{equation}
and
\begin{equation}
\label{rho_mu}
\rho_{\mu}\left(T,\mu\right) = -\frac{1}{L}\left(\frac{\partial\Omega^{*}}{\partial\mu}\right)_{T,V}.
\end{equation}
In order to verify the exact calculations based on the grand canonical ensemble, we performed MC simulations in
the canonical ensemble with the
sampling procedure based on the Metropolis algorithm [35]. The sampling is made with two kind of MC steps:
(i) the exchange of lipid and water molecules positions (ii) the change of lipid molecule
orientation. The specific heat per  lattice site is computed using the fluctuation formula:
\begin{equation}
c_v = \frac{1}{L k_B T^2} \left[ \langle H^2 \rangle - \langle H \rangle^2 \right]
\end{equation}
where here $\langle ... \rangle$ is the average over the microstates in the canonical ensemble.
The simulations were performed for $L \le 4800$. We verified that  the finite size effects are negligible for $L = 4800$
(see Fig.\ref{cv22}a). 

In order to compare thermal properties of this model and the SALR model of Ref.\cite{pekalski:13:0}, we should note that
in the case of the colloidal self-assembly the specific heat was calculated per colloid particle, and the solvent 
was disregarded. The specific heat per unit volume calculated here is a different quantity, therefore in Fig.\ref{cv22}b 
we present $c_v/\rho_s$ representing the specific heat per amphiphilic molecule. 
Note that in this model the specific heat of the pure solvent
(i.e. for $\rho_s=0$) 
 vanishes, since the energy does not fluctuate when all the sites are occupied by the solvent.
In this respect the solvent is analogous to the disregarded solvent in the SALR model. Thus, to compare the thermal 
effects of the amphiphilic and the colloidal self-assembly, we shall compare $c_v/\rho_s$ obtained here
and  the specific heat per
colloid particle calculated in Ref.\cite{pekalski:13:0}.

Let us first discuss the specific heat for $c=g=2$. For such interactions three phases are present on the GS: the solvent-rich
for $\rho_s=0$, the periodic array of bilayers for $\rho_s=2/3$ and the amphiphile-rich for $\rho_s=1$.
The formation of the periodic phase at $T^*=0$  is indicated by the peak for $\rho_s=2/3$ that becomes narrower 
for decreasing temperature 
(Fig.\ref{cv22}a). In Fig. \ref{cv22}b an increase of  $c_v/\rho_s$ for $\rho_s\to 0$ can be observed for the range of $\rho_s$ 
that increases with increasing $T^*$.
This behaviour may be associated with an equilibrium between bilayers and isolated amphiphilic molecules.
In the SALR system the qualitative behavior of the specific heat is similar. The peak for the density corresponding to 
the stability of the periodic phase that becomes narrower for decreasing temperature, and another maximum for a very
small density are both present \cite{pekalski:13:0}.

For $c=1$, $g=2$ only the solvent-rich and the amphiphile-rich phases are stable at the GS. The absence of the periodic phase
leads to a lack of the peak in the specific heat for $\rho_s=2/3$. As in the previous case
an increase of  $c_v/\rho_s$ for $\rho_s\to 0$ can be observed. The $c_v/\rho_s$ in this case is also similar 
to the  specific heat
in the SALR system for the interactions such that only the two homogeneous phases are present on the GS \cite{pekalski:13:0}.

\begin{figure}[th]
\includegraphics[scale=0.9] {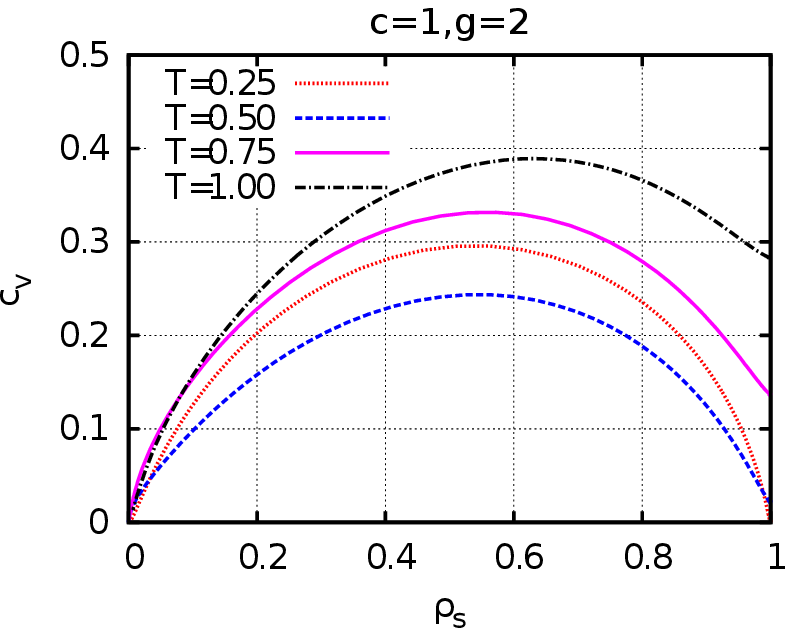}
\includegraphics[scale=0.9] {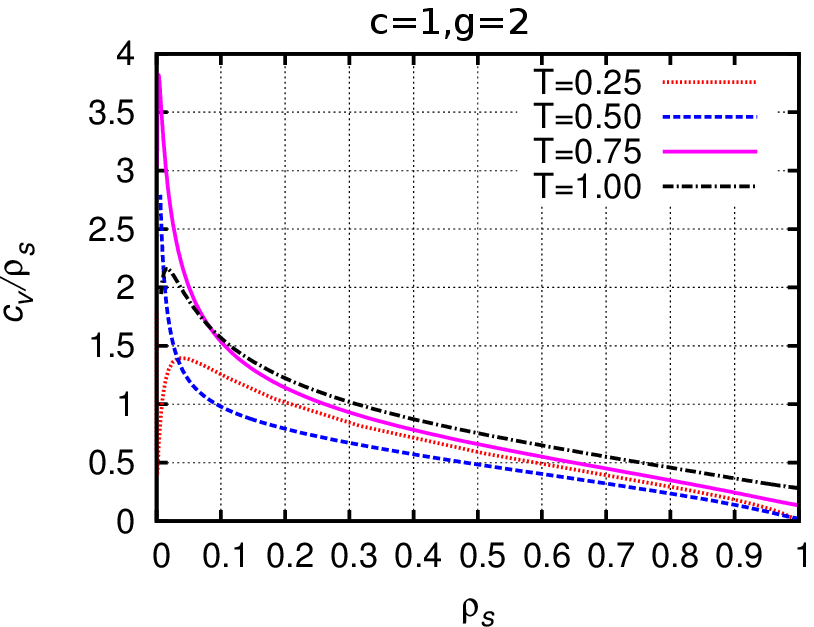}
\caption{  The specific heat for fixed concentration  $\rho_s$ as a function of  $\rho_s$
for $c=1$ and $g=2$ and  different temperatures.
 Left: per unit volume ($c_v$). Right:  per amphiphilic molecule ($c_v/\rho_s$).
Lines represent the 
 analytical results  in thermodynamic limit (Eqs. (\ref{cmu})- (\ref{rho_mu}) and (\ref{Omega})).
}
\label{cv21}
\end{figure}

Note that from Figs.\ref{cv22} and \ref{cv21} it follows that the dependence of $c_v$ on $T$ for fixed $\rho_s$ is nontrivial, 
and differs qualitatively from $c_v(T)$ 
in the lattice gas or Ising model. In the latter models  a single maximum of $c_v(T)$ occurs, whereas in this model two maxima
separated by a minimum are present. The simulation snapshots indicate  formation of solvent-separated bilayers at low $T^*$ for
$\rho_s\le 1/3$. Between the two maxima of $c_v(T)$
 amphiphiles, oriented head-to-head or tail-to-tail, form larger domains separated by domains of solvent. 
At the second maximum random positions and 
orientations of the amphiphiles appear. The two maxima of $c_v(T)$ are present even for $c,g$ such that only the pure solvent and 
pure amphiphiles are present at the GS. In such a case phase-separated solvent and amphiphiles, or random
positions and orientations are present for low $T^*$ or for high $T^*$ respectively, whereas  between the two maxima of
$c_v(T)$ domains of solvent
and properly oriented amphiphiles are formed.

\section{Comparison between the amphiphilic and colloidal self-assembly and concluding remarks}
We have solved exactly the simple lattice model for amphiphilic mixtures  introduced in Ref.\cite{ciach:14:3}. 
The ground-state shows that the model predicts the key properties of aqueous solutions of amphiphilic molecules such as lipids.
It also helps to understand the relation between the degeneracy of the ground state and the low surface tension.

For $T^*>0$ we have obtained oscillatory decay of correlations for the range of $\mu$
corresponding to the GS stability of the periodic phase.
The oscillatory decay of correlations indicates  alternating layers
of the solvent and the properly oriented amphiphilic molecules. 
For small $T^*$ and for the range of the chemical potential $\mu$ 
corresponding to the GS stability of the periodic phase the correlation length
is very large, a few orders of magnitude larger than the molecular size. Thus, the correlation function is consistent 
with a quasi long-range order.
 
The periodic ordering at low temperatures is reflected 
in the shape of the $\rho_s(p^*)$ line
too. The average density $\rho_s\approx 2/3$ is nearly independent of pressure
for a large pressure interval. These two results
are consistent with a formation of solvent-separated bilayers in domains of a size of order of micrometers. 
For increasing $T^*$ the correlation length decreases, and the line  $\rho_s(p^*)$ becomes smoother.  
Finally, the specific heat assumes a maximum for the concentration of amphiphiles $\rho_s=2/3$
corresponding to the GS stability of the periodic phase. The maximum becomes very narrow for low $T^*$. When the periodic phase
is not present on  the GS (see Fig.2), the maximum of the specific heat per amphiphilic molecule for  $\rho_s=2/3$ is absent.

We have found strong similarity of the $T=0$ phase diagrams for the present model for amphiphilic self-assembly (Fig.2)
and for the lattice model for colloidal self-assembly (Fig.1 in Ref.~\cite{pekalski:13:0}).
The phases with oscillatory 
density in the SALR systems or  oscillatory concentration in the amphiphilic mixtures
occur when the repulsion in the case of the colloids or attraction between 
the solvent and  properly oriented amphiphilic molecules  are sufficiently strong.
When the above interactions are weak, 
only two phases are present in the ground state, namely the gas and liquid in the SALR case, and 
the pure water and amphiphile  phases in the present model. 

 It is interesting that the ground state has the same kind of degeneracy at the phase coexistence between the 
periodic phase  and the pure solvent for both, the amphiphilic and the colloidal self-assembly~\cite{pekalski:13:0,pekalski:14:0}. 
In both models the ground state is strongly degenerate, and the surface tension between
the homogeneous and the periodic phase vanishes for $T=0$, 
although  in the case of the colloidal self-assembly the particles  have neither shape nor interaction anisotropy.
Arbitrary number of arbitrarily small droplets of the coexisting phases 
can be present at the phase coexistence. This degeneracy of the ground state means that the macroscopic separation 
of the two phases at $T\to 0$ is not possible, since the formation of an interface does not lead to any
increase of the grand potential. At the same time, because of the microscopic size of the droplets,
one can interpret the degenerate ground state as a disordered phase. 
The region of the $T=0$ phase diagram 
corresponding to the stability of this phase is of zero measure, however, in contrast to 
the remaining, ordinary phases. 
Note that the ultra-low surface tension is a generic property of  systems with competing tendencies in the interactions
that lead to stability of periodic structures, and is not limited to 
amphiphilic  molecules.

The exact results 
for the present model and for the model of colloidal self-assembly show that the low surface tension is a 
more general property of systems with competing interactions, and is not limited to 
amphiphilic  molecules. This  
  confirms the observation of universality of the periodic ordering on the mesoscopic
 length scale that was derived under the assumption of weak ordering \cite{ciach:13:0}.

 The effect of the inhomogeneous density or concentration in the above systems on the equation of state
has been studied in the framework of the statistical thermodynamics only very recently~\cite{pekalski:13:0,ciach:12:0}.
We have found strong similarity between the 
 pressure-density isotherm in the SALR system~\cite{pekalski:13:0} and the pressure-concentration isotherm
 in our model of the amphiphilic mixture.
 The characteristic feature of these lines is the plateau in the density or concentration 
 as a function of  pressure. Similar shapes have also the density - chemical potential and the 
 concentration - chemical potential  lines.
 The plateau occurs when the density or concentration takes the value corresponding to the 
 periodic distribution of the clusters or the bilayers. For the corresponding range  of $\mu$ and $T^*$ the
 correlation functions in both models exhibit exponentially damped oscillatory decay with a very large correlation length.
 The thermal properties are also similar. In both models the specific heat assumes a maximum for the density or concentration
 corresponding to the GS periodic phase. In addition, the specific heat per particle in the SALR case or per amphiphilic molecule 
 in this model increase for decreasing number of particles or amphiphilic molecules. This behavior is associated with
 the equilibrium between the clusters and isolated particles or between the bilayers (or micelles) and isolated amphiphiles.

The main difference between the self-assembly in the amphiphilic 
and colloidal systems concerns the periodic ordering of the pure amphiphiles into the lamellar 
structure that is absent in the dense phase in the colloidal system.

To conclude, the exact results in the present model for amphiphilic systems
and in the lattice model for the SALR systems~\cite{pekalski:13:0} demonstrate close
similarity between different types of self-assembly. 



{\bf Acknowledgments} 

This work is dedicated to Prof. Tomas  Boublik and Prof. Ivo Nezbeda.
A part of this work  was realized within the International PhD Projects
Programme of the Foundation for Polish Science, cofinanced from
European Regional Development Fund within
Innovative Economy Operational Programme "Grants for innovation". 
AC and PR acknowledge the financial support by the National Science Center grant 2012/05/B/ST3/03302.
JP acknowledges the financial support by the National Science Center under Contract Decision No. DEC-2013/09/N/ST3/02551.


\end{document}